# Metasurface-assisted balanced-injection synchronization for turbulence-resilient long-haul chaotic free-space link

Yiqun Zhang[1,2,3†], Mingfeng Xu[1,3,4†], Ning Jiang[2†], Mengjie Zhou[5], Yuhan Zheng[1,3], Sichao Chen[5], Jiazheng Ding[5], Shuangcheng Chen[5], Yong Yu[1,3,4], Xianglei Yan[1,3], Fei Zhang[1,3,4], Yinghui Guo[1,3,4], Mingbo Pu[1,3,4*], Kun Qiu[2*], and Xiangang Luo[1,4*]

[1]State Key Laboratory of Optical Field Manipulation Science and Technology, Institute of Optics and Electronics, Chinese Academy of Sciences, Chengdu 610209, China.

[2]School of Information and Communication Engineering, University of Electronic Science and Technology of China, Chengdu 611731, China.

[3]Research Center on Vector Optical Fields, Institute of Optics and Electronics, Chinese Academy of Sciences, Chengdu 610209, China.

[4]College of Materials Science and Opto-Electronic Technology, University of Chinese Academy of Sciences, Beijing 100049, China.

[5]Tianfu Xinglong Lake Laboratory, Chengdu 610299, China.

†These authors contributed equally to this work.

*Email: pmb@ioe.ac.cn; kqiu@uestc.edu.cn; lxg@ioe.ac.cn

Optical chaotic synchronization between coupled nonlinear lasers underpins most chaos-based applications, including complex laser network dynamics, secure communication, key distribution, and reinforcement learning. In free-space links, however, chaotic synchronization is highly vulnerable to stochastic fluctuation induced by atmospheric turbulence, which results in temporal injection imbalances at symmetric receivers and triggers intermittent desynchronization. Here, we introduce a full Poincaré vector beam-enabled balanced-injection synchronization (BIS) mechanism, which passively mitigates coupling fluctuations and preserves injection symmetry through a complementary metasurface pair, without requiring any channel estimation or active control. Over a 3.2 km urban link under moderately strong turbulence, BIS suppresses coupling power fluctuations by a factor of 4.6 (from 0.4511 to 0.0975). It eliminates desynchronization events and increases the high-quality synchronization probability from 58.6% to 91.0%. This enables a record-high bit rate-distance product of 720 Gbps·km, reducing communication interruption probability by up to 77% compared to Gaussian beam transmission. Our innovative strategy bridges the gap between nanophotonics and engineering optics, offering a new insight into advancing next-generation LiDAR, secure communication, and integrated sensing and communication systems in turbulent environments.

## Introduction

Optical chaos, arising from nonlinear laser dynamics such as external optical feedback[1-3], optical injection[4,5], and optoelectronic feedback[6,7], offers broadband, high-entropy signals with extreme sensitivity to initial conditions, making it a valuable resource for secure communications[8-12], high-precision LiDAR[13-15], secure key distribution[16-18], photonic reservoir computing[19-22], and laser network decision making[23-25]. Chaotic synchronization underlies most of these applications, with established architectures including unidirectional, mutual, and common-injection schemes[26-30]. Unlike the tractable impairments in fiber links[31,32], the mechanisms by which time-varying atmospheric turbulence, a kind of fluid chaos, affects optical chaotic synchronization in free space remain poorly understood. As a result, most demonstrations of optical chaos in free-space link remain restricted to short, turbulence-free laboratory conditions[10,12,33-36], limiting their practical translation to long-range communication and LiDAR.

Significant advances have been made in turbulence mitigation for free-space links, including adaptive optics (AO), AI-based turbulence awareness, and structured light approaches that exploit tailored light-matter interactions to preserve beam integrity. AO systems correct wavefront distortions via real-time feedback but face limitations under strong turbulence, where complex phase aberrations and deep scintillation exceed correction bandwidth[37,38]. Data-driven signal recovery methods offer computational alternatives but face latency and complexity bottlenecks, limiting real-time applicability in high-speed systems[39,40]. In contrast, structured beams target propagation physics that inherently resist distortion via diversity in spatial and polarization degrees of freedom[37,41-45] and can be elegantly realized using metamaterials[46-51], offering integrable hardware solutions for free-space optical (FSO) systems. Despite these promising capabilities, such a strategy has not been systematically explored in the long-distance FSO scenarios, and a mechanistic framework that directly links turbulence-induced vulnerability to chaotic synchronization stability remains lacking, limiting the ability to design robust chaos-based FSO systems.

Here, we reframe turbulence mitigation as a problem of symmetry stabilization rather than full channel recovery. We introduce balanced-injection synchronization (BIS) mechanism, whereby chaotic synchronization is preserved as long as temporal injection powers remain bounded within the allowable imbalance window. BIS is passively realized using full Poincaré beams (FPBs) generated by a complementary spin-multiplexed metasurface pair, which introduces polarization-spatial diversity and redistributes energy among orthogonal components to suppress fluctuation and balance received injection powers. In a 3.2 km urban FSO testbed, FPBs reduce coupling power fluctuations by a factor of 4.6 and increase the probability of high-quality synchronization from 58.6% to 91.0% relative to conventional Gaussian beams under moderately strong turbulence. Integrated with wavelength-division multiplexing (WDM), the same architecture supports AO-free chaotic FSO communication at 240 Gbps using quadrature phase-shift keying (QPSK) modulation. To our knowledge, this represents the first experimental realization of metasurface-based optical communication under realistic kilometer-scale turbulent conditions. These results establish BIS as a mechanism-level strategy for building chaos-based FSO systems, and reveal a metasurface-enabled passive solution for realizing turbulence-resilient LiDAR and secure optical links.

## Results

### Robust chaotic free-space links via metasurface-assisted balanced-injection-synchronization strategy

Figure 1 depicts the conceptual schematic of a BIS-enabled robust chaotic FSO communication system. At the transmitter (Alice), message signals modulated via on-off keying (OOK), phase-shift keying (PSK), or quadrature amplitude modulation (QAM) undergo multidimensional encryption through a chaotic time-frequency aliasing (CTFA) module. This module comprises two dispersive elements and an intermediate phase modulator (PM). The initial dispersive element temporally stretches the optical signal, after which the PM, driven by a chaotic signal ($I_A$) photodetected from a chaotic laser, introduces spectral broadening and chaotic phase scrambling. The subsequent dispersive element converts the spectrally broadened, phase-chaotic waveform into an intensity-chaotic signal, resulting in strong temporal perturbations. Thus, combined dispersion and chaotic phase modulation effectively encrypt the message in temporal, spectral, and phase domains. More details on the encryption process of the CTFA module are presented in Supplementary Note 1.

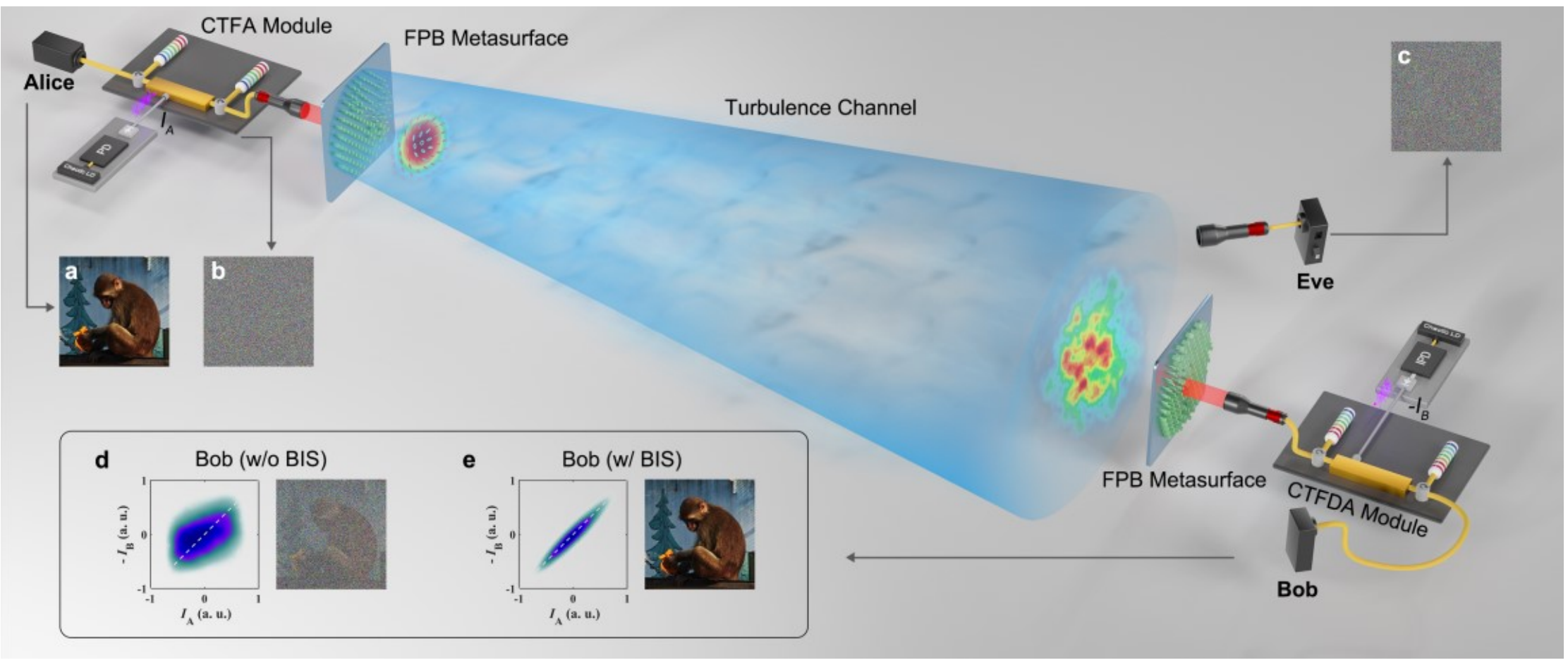

**Fig. 1. Turbulence-resilient chaotic FSO communication scheme based on BIS strategy. a** Original image information. **b** Encrypted image information after the CTFA module. **c** Eavesdropped image information at Eve's side without the CTFDA module. **d-e** Cross-correlation diagrams of the local dynamic chaotic drive signals ($I_A$ and $I_B$) at the transceiver and image information at Bob's side without/with BIS. The color photograph of the monkey was taken by S.C.C., Tianfu Xinglong Lake Laboratory.

The encrypted optical field is then transformed into a lemon-type FPB via a specifically designed FPB metasurface, enabling implementation of the BIS strategy. After long-distance free-space propagation, beam divergence enables potential reception by both authorized (Bob) and unauthorized (Eve) receivers. At Bob's side, a complementary FPB metasurface reconverts the turbulence-distorted beam into a Gaussian profile optimized for fiber coupling. Decryption is executed using a chaotic time-frequency de-aliasing (CTFDA) module, structurally analogous to the CTFA module but driven by an inverse chaotic signal ($-I_B$). Complete decryption requires matching dispersion values (with opposite signs) at corresponding positions, along with isochronal chaotic synchronization between $I_A$ and $I_B$, characterized by high synchronization quality and picosecond-scale synchronization delay.

To better explain the working principle of the robust chaotic FSO communication system, Figure 1a-e illustrates the image transmission performance of the proposed FSO system under varying conditions. Figure 1a,b shows the original and encrypted images at the transmitter, while Fig. 1c confirms that decryption is infeasible without the corresponding CTFDA module. Figure 1d,e compares cross-correlation diagrams and recovered images under turbulence, without and with BIS. Without BIS (i.e., Gaussian beam propagation without applying FPB metasurfaces at the transceiver side), turbulence-induced stochastic fluctuation disrupts synchronization, yielding dispersed correlation scatter and poor image quality. In contrast, BIS, realized passively via FPB metasurfaces, markedly improves synchronization performance, yielding tightly clustered scatter plots along the 45-degree diagonal and substantially enhanced image quality.

**Experimental setup of the metasurface-enabled long-haul FSO communication system**

We experimentally demonstrate a WDM chaotic FSO communication using the setup illustrated in Fig. 2 (see Methods for detailed information). The transmitter and receiver are co-located, while the field trial is conducted over a 3.2 km terrestrial FSO link between two buildings in Xinglong Lake, Chengdu, China. The link is established using 30-element hollow retroreflector arrays positioned 1.6 km away, which reflect the incident beam at its original angle to form a double-pass propagation path.

In previous work, we demonstrated that first-order cylindrical vector beams can mitigate intensity scintillation and enhance the transmission fidelity of optical chaotic signals in indoor turbulence-simulated channels[33]. Here, we employ FPBs for the first time in a field-deployed system, achieving significant improvement in chaotic synchronization performance under real-world turbulence conditions. This type of beam can be decomposed into a superposition of orthogonally polarized transverse spatial modes, $E_1$ and $E_2$, as follows[37,42]:

$$\mathbf{E}(r,\varphi,z) = E_1(r,\varphi,z)\mathbf{e}_1 + E_2(r,\varphi,z)\mathbf{e}_2, \quad (1)$$

where $\mathbf{e}_1$ and $\mathbf{e}_2$ are orthogonal unit vectors and $r$, $\varphi$, $z$ are the cylindrical coordinates.

Here, $\mathbf{e}_1$ and $\mathbf{e}_2$ are taken as the left-handed circular polarization (LCP, $\mathbf{e}_L$) and right-handed circular polarization (RCP, $\mathbf{e}_R$), respectively. The Laguerre-Gauss (LG) basis $LG_{p,l}$ is adopted as the spatial transverse mode, where the first subscript denotes the radial index and the second denotes the azimuthal index. If $E_1$ and $E_2$ carry a zero and a nonzero orbital angular momentum (OAM) value, respectively, the resulting FPB exhibits spatially varying polarization in both angular and radial coordinates, covering all polarization states on the Poincaré sphere, as shown in Fig. 3a.

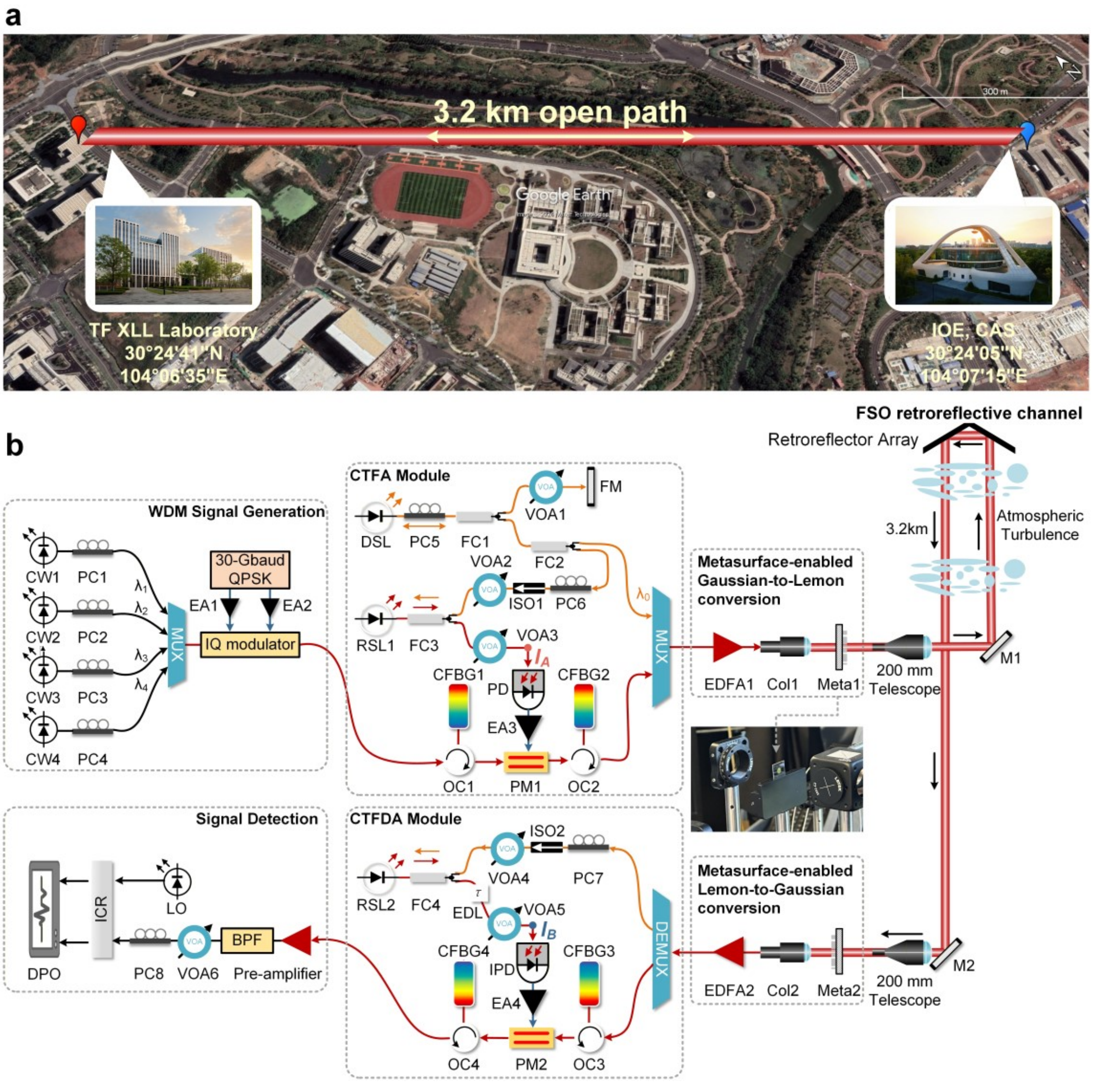

**Fig. 2. Experimental setup of the robust chaotic FSO communication system. a** Overview of the 3.2 km free-space link. Map data: Google Earth. **b** Main configuration and setup. CW, continuous wave; PC, polarization controller; EA, electrical amplifier; DSL, drive semiconductor laser; FC, fiber coupler; VOA, variable optical attenuator; FM, fiber mirror; ISO, optical isolator; RSL, response semiconductor laser; (I)PD, (inverse) photodetector; OC, optical circulator; CFBG,

chirped fiber Bragg grating; PM, phase modulator; EDFA, erbium-doped fiber amplifier; Col, optical collimator; M, mirror; EDL, electrically controlled fiber delay line; BPF, band-pass filter; LO, local oscillator; ICR, integrated coherent receiver; DPO, digital phosphor oscilloscope.

In this work, we employ metasurfaces, a kind of two-dimensional engineered metamaterial comprising subwavelength nanostructures[52-54], to efficiently modulate and demodulate turbulence-resilient vector beams. Two FPB metasurfaces (Meta1 and Meta2) are designed for the system. Meta1, positioned at the transmitter, generates a lemon-type FPB (FPB1) using $LG_{0,0}(r, \varphi, z)\mathbf{e}_L$ and $LG_{0,+1}(r, \varphi, z)\mathbf{e}_R$ modes. Meta2, placed at the receiver, employs the topological charge superposition strategy for beam demodulation (FPB2), using $LG_{0,0}(r, \varphi, z)\mathbf{e}_L$ and $LG_{0,-1}(r, \varphi, z)\mathbf{e}_R$. Through inverse phase modulation, Meta2 cancels the topological charge of the RCP component in the incoming lemon-type FPB, converting this beam back to the fundamental Gaussian mode for subsequent detection. More details about the design method of FPB metasurfaces can be seen in Supplementary Note 2. The fabricated metasurface is shown in Fig. 3b. Each unit cell consists of stacked silicon nanopillars (height $h$ = 800 nm) on a sapphire substrate with a period of $P$ = 700 nm. Structures with optimal transmittance and precise phase matching are selected to construct the FPB metasurfaces (Supplementary Fig. S2), each with a footprint of 5 mm × 5 mm. Figure 3c presents the scanning electron microscopy (SEM) image of the manufactured Meta1, revealing high verticality and uniformity.

To verify the metasurfaces' capability in generating FPBs, we measure the intensity and polarization distributions of FPB1 and FPB2, as well as the intensity profiles of their LCP and RCP components, as depicted in Fig. 3d,e. Transmission efficiency and polarization conversion efficiency (PCE) of both metasurfaces are subsequently calculated (see Supplementary Note 2 for more details). The generated beams exhibit Gaussian-like intensity profiles, and the measured polarization distributions closely align with simulation results. Meta1 achieves a transmission efficiency of 87.8% and a PCE of 83.5%, while Meta2 yields 84.6% and 87.8%, respectively. These results confirm the high precision and effectiveness of the designed metasurfaces in generating desired FPBs.

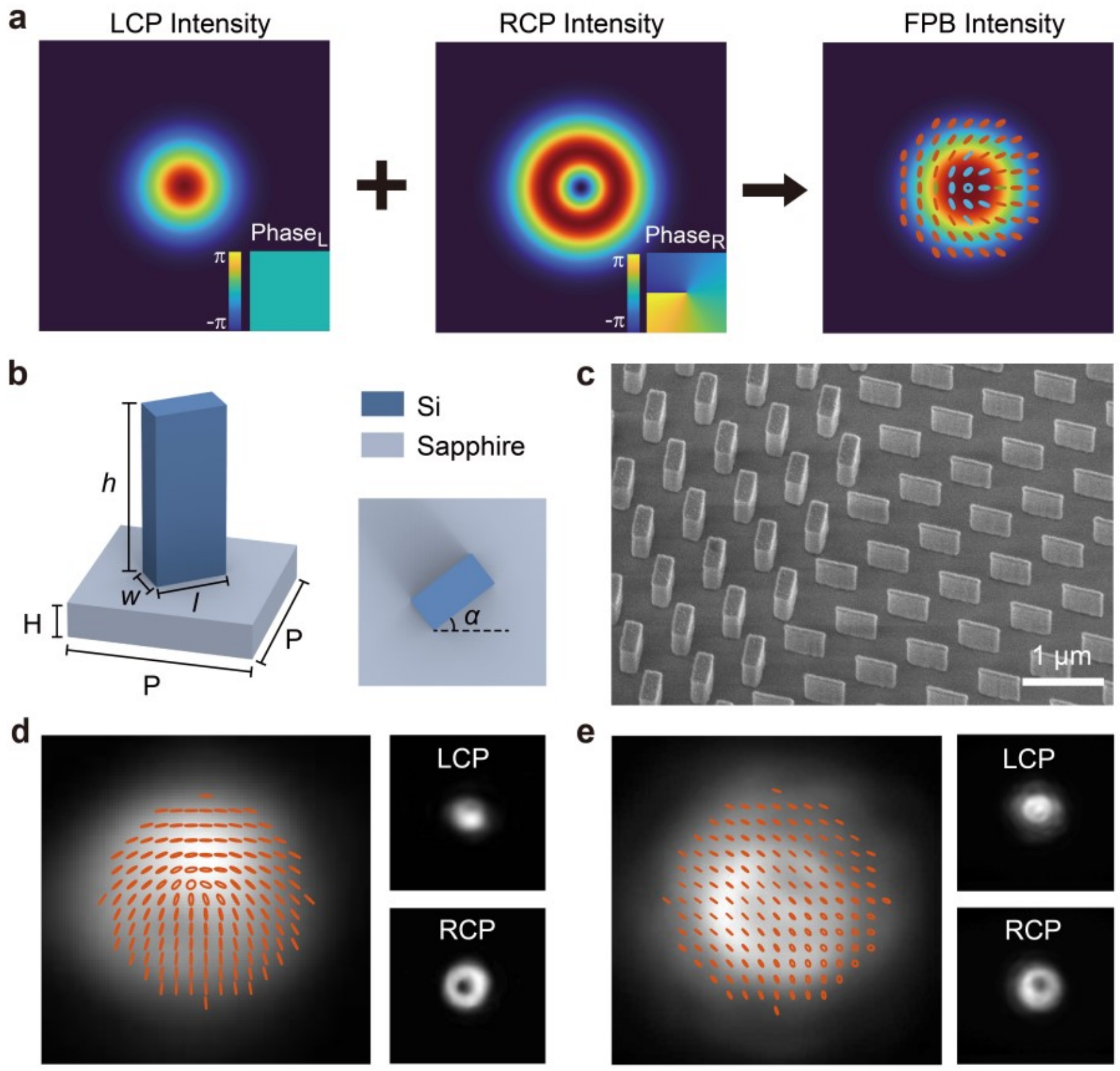

**Fig. 3. Spin-multiplexed metasurfaces design and their performance. a** A general FPB generation method. LCP light carries the phase of $LG_{0,0}$, and the RCP light carries the phase of $LG_{0,1}$, which is superimposed to produce a lemon-type FPB. **b** Left: schematic of the meta-atom made up of silicon in the FPB metasurface. Right: a top view of the meta-atom, where the width $w$, length $l$, and orientation $\alpha$ are varied for different meta-atoms. **c** Scanning electron microscopy image of Meta1. The SEM image of Meta2 shows similar structural characteristics to Meta1. **d** Left: measured intensity and polarization distributions of FPB1. Right: experimentally obtained LCP intensity and RCP intensity of FPB1. **e** Left: measured intensity and polarization distributions of FPB2. Right: experimentally obtained LCP intensity and RCP intensity of FPB2.

## Balanced-injection-synchronization empowered by FPB metasurfaces

The disruptive effect of atmospheric turbulence on common-injection-induced chaotic synchronization primarily stems from intensity scintillation, which drives stochastic fluctuations in the coupled optical power (COP) and induces temporal injection power imbalance that destabilizes the synchronization manifold. Within the BIS framework, synchronization locking is preserved only when this instantaneous power asymmetry remains bounded within an allowable imbalance window. Figure 4a quantifies this bound in a back-to-back configuration, taking a cross-correlation coefficient (CC) of 0.9 as the criterion for high-quality synchronization[20,55]. As injection mismatch increases, CC decreases monotonically, falling below 0.9 once the mismatch exceeds approximately 1.5 dB, thus defining the allowable imbalance window. Phase-space evidence is provided in Fig. 4b, where three-dimensional trajectories of chaotic signals $I_A$ and $I_B$ are reconstructed from measured time series[56]. At 0.5 dB mismatch, the reconstructed attractors are topologically congruent (CC = 0.94), whereas at 3dB, the attractors diverge and CC drops to 0.78, indicating manifold rupture and transition to an unsynchronized regime. Together, Figs. 4a,b operationalize the BIS principle: high-quality synchronization is preserved when COP fluctuations remain within ~1.5 dB, and turbulence countermeasures should therefore be evaluated by their ability to suppress balance-breaking fluctuations within this bound.

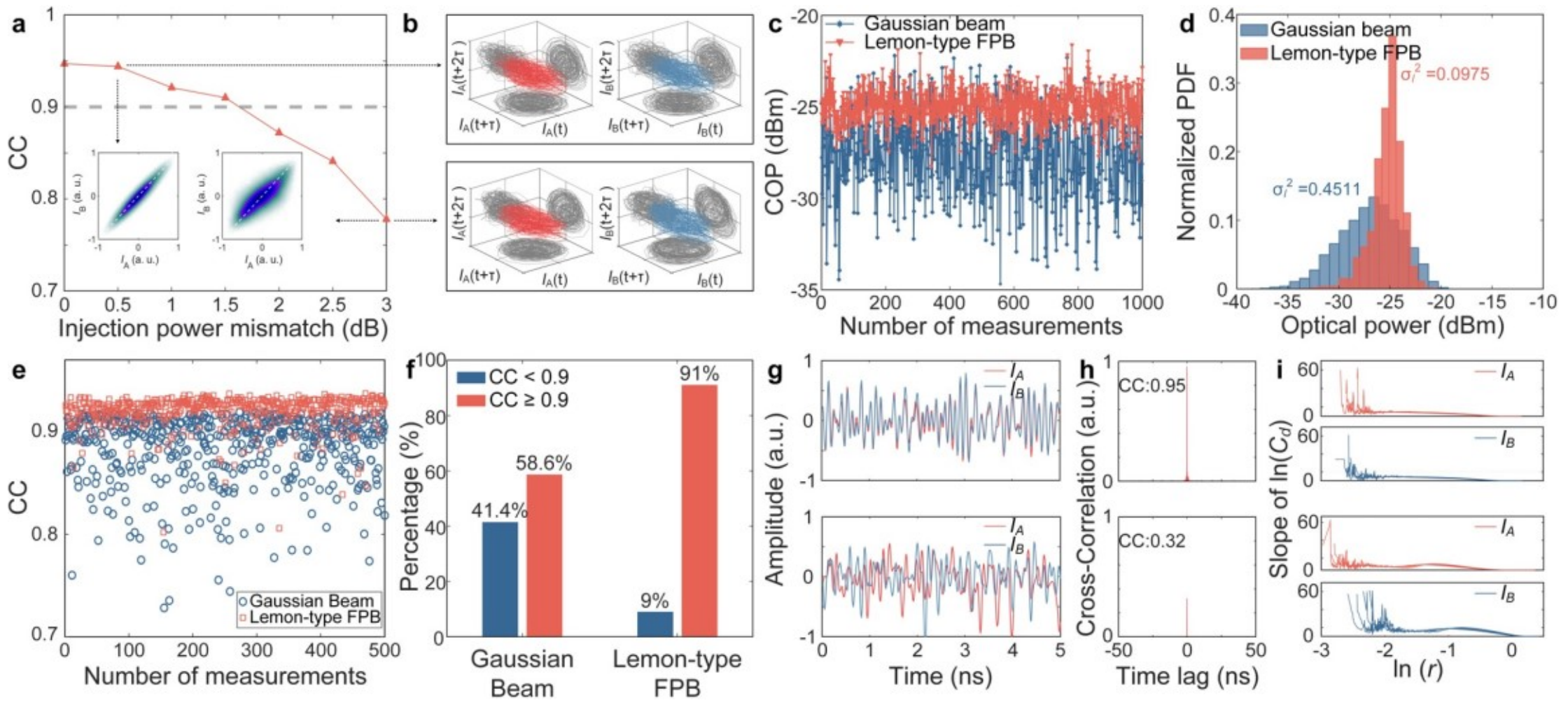

**Fig. 4. Field trial results of BIS performance for Gaussian beam and lemon-type FPB. a** Effect of injection power mismatch on chaotic synchronization for two chaotic lasers at the transceiver under the back-to-back case. Insets show two-dimensional cross-correlation function curves at mismatches of 0.5 dB and 3 dB. **b** Three-dimensional phase space trajectories of chaotic signals $I_A$ (red) and $I_B$ (blue) under mismatches of 0.5 dB (top) and 3 dB (bottom). **c** Measured COP over time for the lemon-type FPB (red) and Gaussian beam (blue) under a 3.2 km terrestrial FSO link. **d** Probability density function distributions of the COP for both beams. **e** Measured chaotic synchronization performance (CC) for Gaussian beam and lemon-type FPB over 500 consecutive trials. **f** Percentage of high-quality chaotic synchronization (CC ≥ 0.9) for both beams. **g** Time-domain waveforms of $I_A$ and $I_B$ for two examples selected from 500 measurements, corresponding

to CC values of 0.95 (top) and 0.32 (bottom). **h** One-dimensional cross-correlation functions between $I_A$ and $I_B$ for these two examples. **i** Slope of the correlation integral $C_d(r)$ versus the sphere radius $r$ for the signals $I_A$ and $I_B$ in the same two examples.

To experimentally validate the BIS performance of FPB metasurfaces, a conventional Gaussian beam is chosen as a control. Due to experimental constraints, simultaneous measurements of both beams are not feasible. To ensure fair comparison, atmospheric turbulence strength is continuously monitored in real-time using a Differential Image Motion Monitor (DIMM), and only data collected under comparable turbulence conditions are selected for analysis. Further details on turbulence characterization are provided in Methods section. A 90:10 fiber coupler is connected after Col2 at the receiver, with 10% output routed to a power meter for real-time monitoring of the COP. Figure 4c illustrates the COP fluctuations of the Gaussian beam and lemon-type FPB measured over a 1000-second observation period under moderately strong turbulence conditions (average refractive index structure parameter $C_n^2 \sim 4\times10^{-14}$ $m^{-2/3}$, details are shown in Supplementary Fig. S4). The lemon-type FPB exhibits reduced power fluctuations, with a narrower dynamic range of ±3 dB and a standard deviation of 1.51 dB, in contrast to ±6 dB and 3.04 dB for the Gaussian beam, respectively. Figure 4d presents the corresponding probability density function (PDF) distributions of the COP, further confirming that the lemon-type FPB exhibits higher and more concentrated COP. The measured scintillation index $\sigma_I^2$, obtained from the standard free-space intensity scintillation formula[37], is 0.0975 for the FPB, representing a 78.38% reduction relative to the Gaussian beam's value of 0.4511, corresponding to a 4.6-fold enhancement in scintillation suppression.

Figure 4e illustrates the chaotic synchronization performance, based on 500 consecutive measurements for each beam under similar 3.2 km atmospheric turbulence conditions. The lemon-type FPB (red squares) exhibits a broader concentration of CC values above 0.9, whereas the Gaussian beam (blue circles) shows more dispersed and fluctuating CC values. To facilitate a clearer comparison, the proportion of measurements with CC $\geq$ 0.9 is calculated for each beam and presented as histograms in Fig. 4f. The probability of achieving high-quality chaotic synchronization is 58.6% for the Gaussian beam, which increases markedly to 91.0% for the lemon-type FPB. This performance enhancement is closely associated with improved COP stability. As shown in Fig. 4c, the lemon-type FPB exhibits a COP fluctuation with a standard deviation of 1.51 dB, aligning with the allowable imbalance window (~1.5 dB). This suggests that FPB effectively suppresses turbulence-induced intensity scintillation, maintaining injected power within the allowable imbalance window and thereby enabling robust, high-quality BIS under realistic atmospheric conditions.

Figure 4g-i illustrates the dynamic chaotic behavior of chaotic signals $I_A$ and $I_B$ for two representative cases (CC = 0.95 and CC = 0.32), selected from 500 measurements shown in Fig. 4e. For the high-quality synchronization case (CC = 0.95), the time series of $I_A$ and $I_B$ closely overlap, and the one-dimensional cross-correlation function exhibits a sharp delta-like peak, indicating strong synchronization. Furthermore, the correlation dimension $d$, which quantifies the complexity of chaotic attractors via phase-space correlation[57], provides additional evidence of synchronization. The slopes of correlation integrals for $I_A$ and $I_B$ display similar trends and converge within comparable radius ranges, reflecting matching dynamical complexities and a synchronized attractor structure. Conversely, in the poor synchronization case (CC = 0.32), the time series significantly diverge, the maximum value of cross-correlation is substantially lowered, and correlation integral slopes show distinct trends and convergence intervals, clearly indicating weak correlation and unsynchronized chaotic dynamics.

**Turbulence-resilient 240 Gbps QPSK FSO transmission**

We demonstrate a 240 Gbps WDM FSO transmission under the same turbulent link, with each wavelength carrying 60 Gbps QPSK data. Figure 5a-d presents the calculated bit error rate (BER) values and corresponding PDF distributions for the Gaussian beam and lemon-type FPB across 50 measurements on each channel. Without metasurfaces-assisted BIS, turbulence-induced scintillation leads to significant synchronization instability, causing over half of the BER values to exceed the hard-decision forward error correction (HD-FEC) limit of $3.8\times10^{-3}$. In contrast, with BIS, BER fluctuations are reduced and primarily fall below the HD-FEC limit. Figure 5e quantifies the proportion of BER values below this limit for each beam across all channels. Under the kilometer-scale terrestrial FSO link, only about 30%-40% of Gaussian beam transmissions meet the HD-FEC limit, whereas this proportion rises to approximately 66%-86% for the lemon-type FPB, corresponding to a 51%-77% reduction in communication interruption probability. In addition, we further evaluated the transmission performance of these two cases by measuring BER as a function of received optical power. For the 3.2 km free-space link, the FPB exhibits a power penalty reduction of approximately 1.8 dB compared to the Gaussian beam (see more details in Supplementary Note 3).

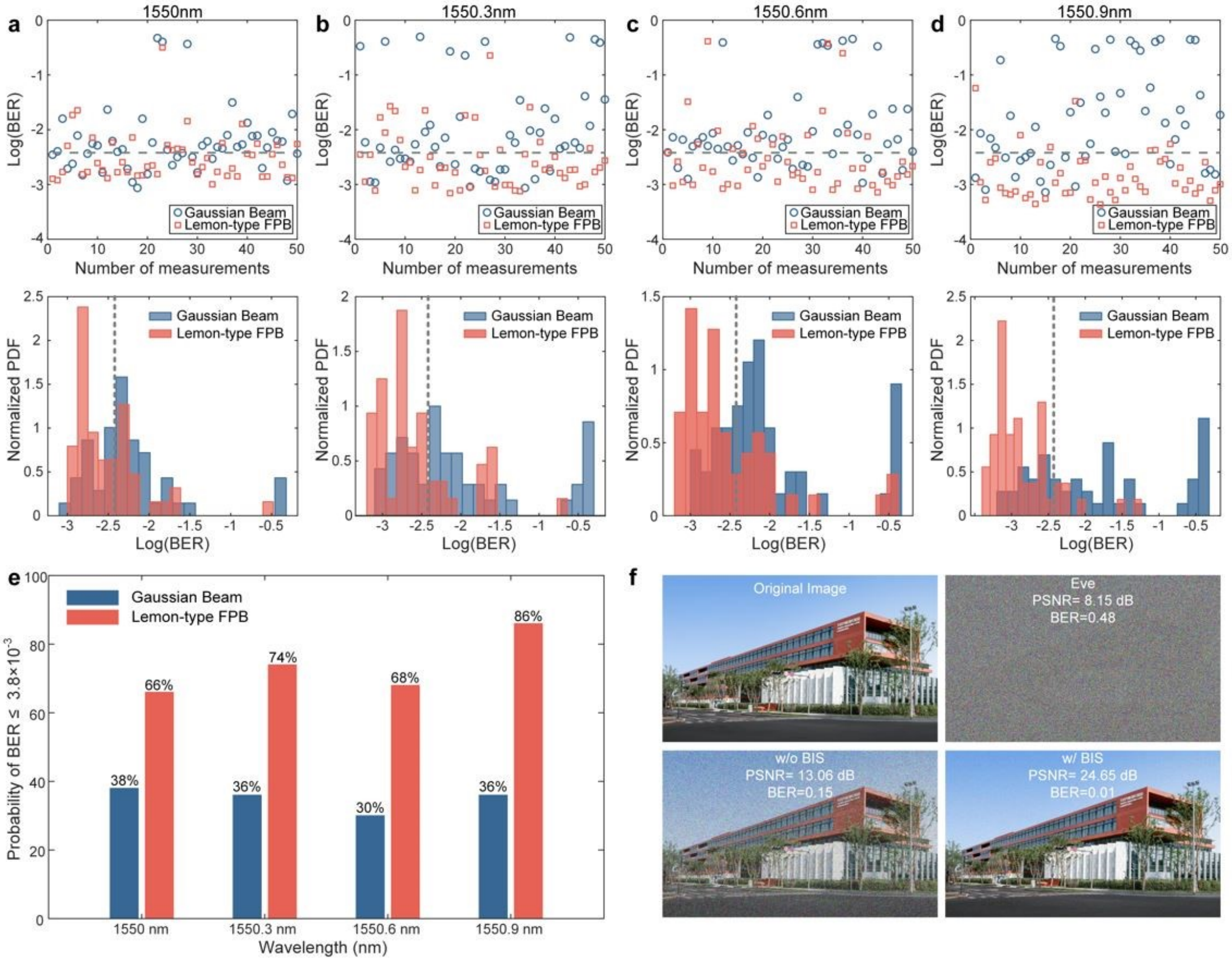

**Fig. 5. Experimental results for 240 Gbps QPSK data transmission in a coherent FSO link under real-world turbulence, with and without FPB metasurfaces. a-d** Measured BER performance and corresponding PDF distributions for the Gaussian beam (blue circles) and lemon-type FPB (red squares) at wavelengths of 1550 nm, 1550.3 nm, 1550.6 nm, and 1550.9 nm, respectively. **e** Histogram of the proportion of BER values below the HD-FEC limit for both beams across all channels. **f** Original and retrieved images under different reception scenarios. PSNR values and BER values are calculated as the mean PSNR and BER across the four channels.

Figure 5f presents the image transmission performance after applying the CTFA

module and metasurfaces-assisted BIS through the turbulent channel. Pixels with incorrectly received information appear as noise. For the eavesdropper lacking the CTFDA decryption module, the peak signal-to-noise ratio (PSNR) is only 8.15 dB, confirming the effectiveness of the CTFA technique in preventing information leakage. For the legitimate receiver, without BIS, synchronization instability leads to degraded image quality, yielding a PSNR of 13.06 dB. In contrast, with FPB metasurfaces employed, the transmission becomes more robust, improving the PSNR to 24.65 dB and reducing the BER by an order of magnitude.

**Discussions**

Here, we have introduced a passive BIS strategy based on a complementary spin-multiplexed metasurface pair to mitigate atmospheric turbulence in chaotic free-space links. By exploiting the inherent interaction between structured light and the turbulent medium, we demonstrate that the generation of lemon-type FPB via metasurfaces effectively suppresses stochastic intensity scintillation, thereby stabilizing injection power symmetry within the allowable imbalance window. From the quantum framework perspective, vectorial structured light transmission in complex media exhibits topological invariance[58-60]. The combined topological protection, arising from OAM diversity and spatial polarization diversity, enables vector beams to maintain high robustness and stability, even as the properties of the medium vary. In contrast to conventional AO systems, which rely on active feedback and the assumption of channel reciprocity, the proposed metasurface-based method offers a passive, reciprocal-free, and low-latency solution that can be symmetrically deployed at both ends of the link to support bidirectional communication. The resulting BIS over kilometer-scale horizontal links highlights the potential of this approach for enhancing the reliability of next-generation chaos-based FSO systems operating in complex atmospheric environments.

As summarized in Table 1, compared to state-of-the-art chaotic FSO communication demonstrations, our work achieves successful 3.2 km free-space transmission of four-channel 240 Gbps QPSK signals under moderately strong atmospheric turbulence. We note that our experimental results were obtained over an urban link with sufficiently large receiving apertures. In more practical long-range scenarios, limited aperture size and beam divergence may cause power loss and mode distortion, which could be mitigated by integrating with multiple-input multiple-output techniques[62] or optical phase conjugation[63]. Beyond kilometer-scale FSO communication, metasurface-assisted BIS strategy also holds promise for advanced applications such as turbulence-resilient chaotic secure key distribution and multifunctional LiDAR systems capable of surface structure and velocity-direction sensing[64].

**Table1. Performance comparison of the state-of-the-art chaotic FSO transmission**

| Ref | Chaotic synchronization mechanism | Spatial mode | Transmission wavelength (nm) | Transmission distance (m) | Transmission capacity (Gbps) | Turbulence-resilient capacity |
|---|---|---|---|---|---|---|
| [61] | Digital chaos-based synchronization | Gaussian Beam | 690 | 5000 | $6\times10^{-5}$ | No |
| [35] | Digital chaos-based synchronization | Gaussian Beam | 1550 | 3 | 10 | No |
| [10] | Unidirectional injection | Gaussian Beam | 5700 | 1 | $5\times10^{-4}$ | Yes |
| [36] | Unidirectional injection | Gaussian Beam | 9300 | 30 | $8\times10^{-3}$ | Yes |
| [34] | Unidirectional injection | Gaussian Beam | 1550 | 10 | 8 | No |
| [12] | Unidirectional injection | OAM Beam | 1550 | 2 | 20 | No |
| [33] | Unidirectional injection | Cylindrical vector Beam | 1550 | 10 | 22 | Yes |
| This work | Common-light-induced BIS | Full-Poincaré beams | 1550 | 3200 | 240 | Yes |

## Methods

### Experimental details of kilometer-scale metasurface-enabled chaotic FSO communications

As shown in Fig. 2, at the transmitter, a 60 Gbps QPSK signal is generated and pulse-shaped by a root-raised cosine filter (0.1 roll-off). The signal is then converted from digital to electrical form using an arbitrary waveform generator (AWG, Keysight M8195A) operating at 64 GSa/s with 20 GHz analog bandwidth. The electrical signal is uploaded to a high-bandwidth coherent driver modulator (NeoPhotonics HB-CDM, Class 40), enabling coherent modulation of a WDM optical signal. The four-wavelength optical carrier is provided by a continuous-wave integrable tunable laser assembly (EXFO, FTBx-2850-1-4-C-H) with a linewidth of 100 kHz and central wavelengths of $\lambda_1$ = 1550.0 nm, $\lambda_2$ = 1550.3 nm, $\lambda_3$ = 1550.6 nm, and $\lambda_4$ = 1550.9 nm.

The WDM signal is then routed to the CTFA module for encryption, which includes the first dispersive element (CFBG1), a PM1 (CONQUER, KG-PM-15-20G-PP-FA) with a half-wave voltage of 3.1 V and 20 GHz bandwidth, and a second dispersive element (CFBG2). CFBG1 is a FBG-based tunable dispersion compensation module (TeraXion, TDCMB-C000-OE-BF01) providing a dispersion of 1600 ps/nm. PM1 is driven by a local chaotic signal $I_A$, generated by a response semiconductor laser (RSL1) under common optical injection. The common chaotic light originates from a conventional optical-feedback-based drive semiconductor laser (DSL) operating at a central wavelength of $\lambda_0$ = 1549.5 nm. After polarization adjustment via polarization controller (PC5), the DSL output is split by a 20:80 fiber coupler (FC1): 20% is fed back through a loop with a variable optical attenuator (VOA1) to control the feedback strength, while the remaining 80% serves as the common chaotic source. This chaotic light is further divided by a 10:90 FC2: 90% is injected into RSL1 to generate the chaotic driving signal $I_A$, and 10% is wavelength-multiplexed with the encrypted WDM signal. CFBG2 is also a tunable dispersion compensation module, providing a dispersion of 1000 ps/nm.

After amplification by EDFA1, the WDM signal is emitted into the vector optical field manipulation module via a collimator Col1 (Thorlabs, TC25APC-1550, beam diameter 4.65 mm). A lemon-type FPB is generated using the custom-designed FPB metasurface (Meta1) and transmitted through a 3.2 km FSO link using a 10× beam-expanding telescope (SMT-JSXT). The FSO link spans an urban area in Chengdu, China, with hollow retroreflector arrays (AR-30) positioned 1.6 km from the transmitter to retroreflect the beam back through the same atmospheric path, where it is collected by the same telescope.

At the receiver, the incoming lemon-type FPB is converted back into a Gaussian beam using a second FPB metasurface (Meta2) and coupled into the single-mode fiber via Col2. The signal is then amplified by EDFA2 and passed through a programmable optical filter (COHERENT, WaveShaper 4000A), which separates the common chaotic light from the WDM signal. The filtered common chaotic light is injected into RSL2 to generate the local chaotic driving signal $I_B$ for driving PM2, while the filtered encrypted

WDM signal is routed to the CTFDA module for decryption. During the generation of local chaotic driving signals by RSL1 and RSL2, PC6 and PC7 are used to adjust the polarization state of the common chaotic light to achieve synchronization between $I_A$ and $I_B$. VOA2 and VOA4 control the injection power into RSL1 and RSL2, while VOA3 and VOA5 regulate the voltage magnitude applied to PM1 and PM2, respectively. In the CTFDA module, CFBG3 and CFBG4 have dispersion values of -1000 ps/nm and -1600 ps/nm, respectively. The decrypted WDM signal is demultiplexed using a variable optical bandpass filter with a 40 GHz bandwidth. The selected channel is then passed through PC8 and fed into an integrated coherent receiver (Fujitsu FIM24706/201). A local oscillator laser, precisely tuned to the target wavelength, enables effective intradyne detection. The detected signal is subsequently captured by a digital storage oscilloscope and processed using offline digital signal processing algorithms.

In our experiment, the DSL and RSLs are distributed-feedback lasers driven by an integrated low-noise driver (OVLINK, OFLD-2000B). The DSL operates at an output power of 8 dBm with a fixed optical feedback strength of -10 dB. The free-running wavelengths of RSL1 and RSL2 are set to 1549.45 nm, with emission and injection powers configured at 0 dBm and -2 dBm, respectively. PDs with 10 GHz bandwidth and differential output are used to ensure that the chaotic driving signals $I_A$ and $I_B$ exhibit opposite amplitudes. Electrical amplifiers provide a gain of 45 dB over a frequency range of 0.3 to 20 GHz. The output power of EDFA1 is fixed at 21 dBm. A high-speed digital oscilloscope (Keysight, UXR0254A) captures the electrical signals at a sampling rate of 128 GSa/s across four 25-GHz bandwidth channels. An optical spectrum analyzer with 0.01 nm resolution is used to record the optical spectra of signals.

Chaotic synchronization between $I_A$ and $I_B$ is critical for successful decryption. Because the CTFA module performs high-speed encryption entirely in the optical domain, synchronization delay must be minimized to ensure temporal alignment with the encrypted signal. For data rates on the order of tens of Gbit/s, synchronization delays must be controlled within tens of picoseconds to avoid decryption failure. In practical deployment, device-level mismatches between the CTFA and CTFDA modules, such as pigtail length differences and circuit-induced delays, make precise synchronization delay identification and compensation particularly challenging. To address this, we propose a self-homodyne phase autocorrelation time-delay detection method, which enables synchronization delay measurement with picosecond-level accuracy (see more details in Supplementary Note 4).

**Fabrication of the FPB metasurfaces**

First, the pre-cleaned silicon-on-sapphire (SOS) substrate was spin-coated with the electron resist (Ma-N2403) onto the Si film at 3000 rpm and baked at 100°C for 60 seconds. The resist was then patterned by electron-beam lithography at a dose of 240 $\mu C/cm^2$ and an accelerating voltage of 125 kV. After immersion in the ma-D525 solution for 30 seconds, a pattern was generated on the surface. Inductively coupled plasma etching (ICP, Sentech SI500) was subsequently employed to etch the structures using $C_4F_8$ and $SF_6$ gas. Finally, the residual resist was removed by reactive ion etching (RIE, Sentech SI500). The schematic of the FPB metasurface fabrication process is shown in Supplementary Fig. S9.

**Turbulence characterization of the 3.2 km real-world free-space link**

To accurately characterize the strength of atmospheric turbulence along the transmission path, a Differential Image Motion Monitor (DIMM) was deployed near the transmitter to measure the Fried parameter ($r_0$) in real time. The DIMM system uses a telescope with a masked aperture that forms two spatially separated subapertures, enabling the calculation of $r_0$ by measuring the variance in differential image motion[65]. Although $r_0$ is wavelength-dependent, measurements taken at 550 nm using a beacon source (placed at the hollow retroreflector array's side) still provide a valid quantitative indicator of atmospheric transmission quality for a laser operating at 1550 nm. In this work, the turbulence strength is described using the atmospheric refractive index structure constant $C_n^2$, which relates to $r_0$ through the following expression[66]:

$$r_0 = \left[ 0.423 k^2 \int_0^L C_n^2(z)\, dz \right]^{-3/5}, \tag{2}$$

where $k$ is the optical wavenumber and $L$ is the total FSO transmission distance (3.2 km in this case).

**Acknowledgements**

This work was supported by the National Natural Science Foundation of China (NSFC) (Grants Nos. 62575286, 62475036, U24A6010, 62401540, and 62175242), China Postdoctoral Science Foundation (2024M753238), and Stability Program of National Key Laboratory of Security Communication (WD202405). The authors would like to thank Zheng Song for his help with experimental measurements.

**Authors' contributions**

Y.Q.Z., M.F.X., and N.J. conceived the idea and planned the research. Y.H.Z. designed and fabricated the metasurfaces. Y.H.Z. and M.J.Z. carried out experiments on metasurface measurements. Y.Q.Z., M.J.Z., S.C.C., and Y.Y. designed the FSO experiments and performed the measurements. Y.Q.Z., S.C.C., J.Z.D., and X.L.Y. contributed to the data analysis. Y.Q.Z., M.F.X. wrote the manuscript. F.Z., Y.H.G., and M.B.P. presented suggestions for improving the quality of this work. M.B.P., K.Q., and X.G.L. supervised the project. All authors contributed to the discussions and manuscript writing.

**Additional information**

Competing financial interests: The authors declare no competing financial interests.

**Data availability**

All the data used to generate the plots and support the findings reported in this study are available from the corresponding authors upon request.

## Supplementary Note 1: Encryption mechanism of the chaotic time-frequency aliasing model

Figure S1 illustrates the detailed schematic of the chaotic time-frequency aliasing (CTFA) encryption process for both intensity and coherent signals. High-speed optical data, carried over multiple wavelength channels, is first multiplexed via WDM and then enters the CTFA module. The encryption module consists of an electro-optical phase modulator (PM) and two dispersive elements. In our work, chirped fiber Bragg gratings (CFBG) are employed as the dispersive elements. Figure S1(a) illustrates the optical spectral evolution of the multichannel signal within the CTFA encryption module:

Initially, CFBG1 introduces group velocity dispersion ($\beta_2$) to differentiate the transmission delay across frequency components, resulting in temporal broadening and inter-symbol interference. Its frequency response is given by:

$$H_{CFBG1}(\omega)=\exp\left(i\frac{1}{2}\beta_2\omega^2\right). \tag{1}$$

The output field $E_1(t)$ after CFBG1 can be expressed as:

$$E_1(t)=F^{-1}\left[F\left(E_0(t)\right)\cdot H_{CFBG1}(\omega)\right], \tag{2}$$

where $F(\cdot)$ and $F^{-1}(\cdot)$ denote the Fourier and inverse Fourier transforms, respectively. This process alters the temporal structure without significantly modifying the signal's spectral shape.

Next, the chaotic driving signal generated by the local RSL drives the PM to impose random phase modulation on $E_1(t)$, producing multi-order spectral broadening. The modulated output $E_2(t)$ is given by:

$$E_2(t)=E_1(t)\cdot\exp\left(i\phi(t)\right), \tag{3}$$

$$\phi(t)=\pi\cdot\frac{V_{chaos}(t)}{V_\pi}, \tag{4}$$

where the phase perturbation term $\phi(t)$ is defined in Eq.(4), with $V_{chaos}$ as the chaotic driving voltage, $V_\pi$ the PM half-wave voltage, and their ratio defining the phase modulation index (PMI). As PMI increases, spectral broadening becomes more pronounced, and adjacent channels increasingly overlap, ultimately forming an ultra-broadband flattened spectrum that obscures both channel spacing and wavelength information, thereby enhancing spectral-domain security.

Finally, the dispersion effect of CFBG2 (frequency response $H_{CFBG2}(\omega)$) converts the phase perturbation into an intensity-like noise signal. The output field $E_3(t)$ after CFBG2 is given by:

$$E_3(t) = F^{-1}\left[F\left(E_2(t)\right)\cdot H_{CFBG2}(\omega)\right]. \tag{5}$$

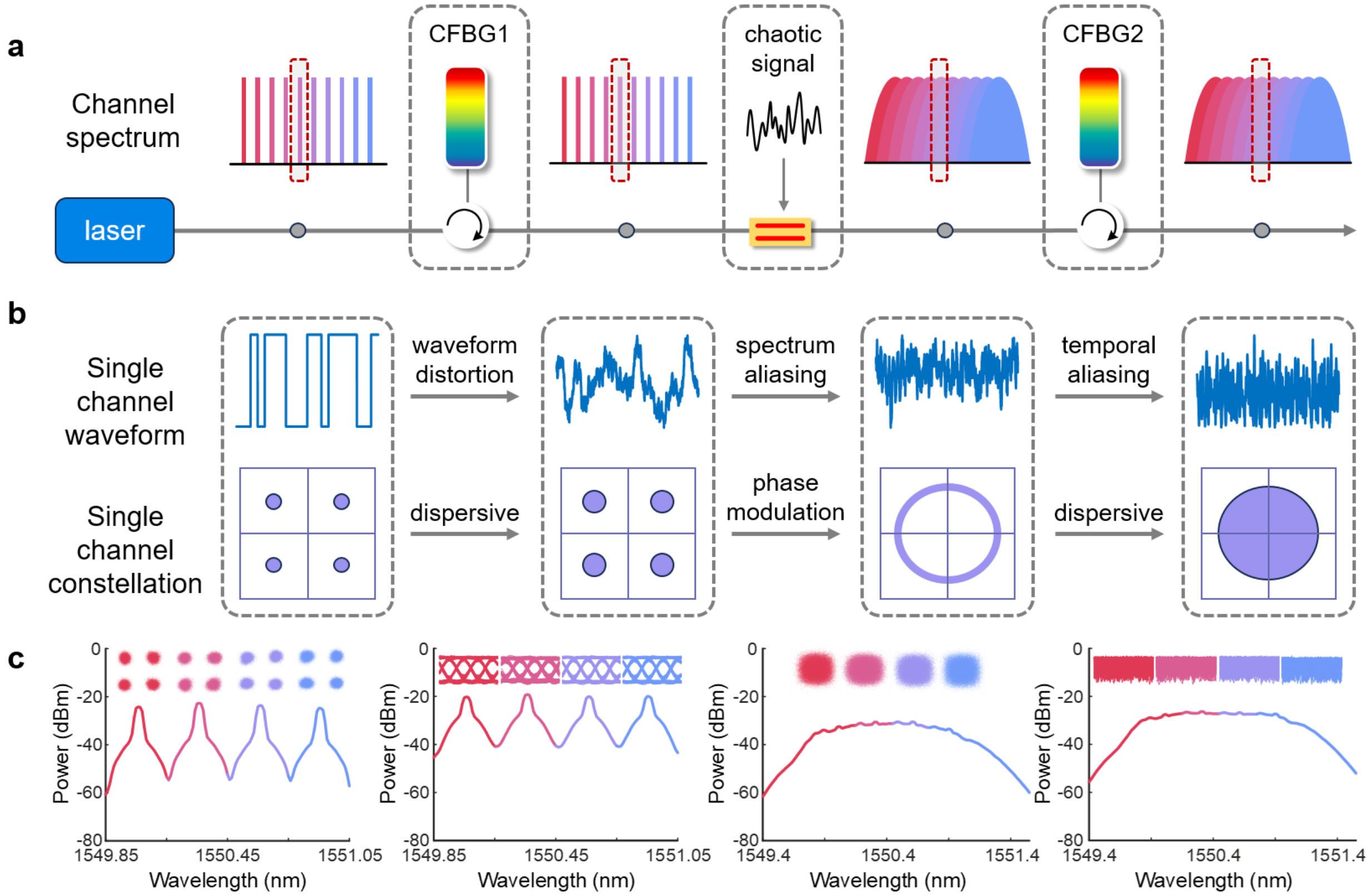

**Fig. S1. CTFA encryption process and experimental validation for multichannel signals.** (a) Schematic illustration of spectral evolution across each stage of the CTFA module for multichannel signals. (b) For a representative channel: top row shows the evolution of the time-domain waveform as the signal propagates through each component of the CTFA module; bottom row displays corresponding changes in QPSK constellation diagrams. (c) Experimental results: from left to right—original spectra and QPSK constellation diagrams for four channels, original spectra and NRZ-OOK eye diagrams for four channels, encrypted spectra and QPSK constellations, and encrypted spectra and OOK eye diagrams.

Figure S1(b) illustrates the encryption evolution for both intensity-modulated (e.g., NRZ-OOK) and coherently modulated (e.g., QPSK) signals:

(1) NRZ-OOK signal: Once sufficient dispersion is introduced by CFBG1, inter-symbol interference and temporal aliasing are induced. The chaotic phase modulation from the PM introduces spectral aliasing, further exacerbating time-domain overlap. CFBG2 then transforms this multidimensional perturbation into a time–frequency–phase-aliased noise-like signal via phase-to-intensity conversion.

(2) QPSK signal: Chaotic phase modulation causes the coherent signal's constellation points to spread into a ring-shaped distribution. For multichannel QPSK, spectral aliasing intensifies this spreading in the IQ plane. After CFBG2-induced phase-intensity conversion, the constellation becomes fully dispersed across the IQ plane, forming a disordered distribution that deeply masks the signal structure and modulation format.

Figure S1(c) presents the experimental validation of the CTFA encryption effect for four-channel QPSK (60 Gbps/channel) and NRZ-OOK (20 Gbps/channel) signals:

(1) Before encryption: The optical spectrum clearly shows distinct channel spacings at 1550.0 nm, 1550.3 nm, 1550.6 nm, and 1550.9 nm. The QPSK constellation diagrams exhibit well-defined quadrants with an error vector magnitude (EVM) of approximately 17%, and the NRZ-OOK eye diagrams display clear openings, indicating high signal fidelity.

(2) After encryption: The optical spectrum is spectrally broadened into a fully blended continuum, eliminating identifiable wavelength features. The QPSK constellation diagrams become highly disordered and diffuse (EVM ≈ 40%), while the NRZ-OOK eye diagrams are fully closed, indicative of deep temporal and modulation-format masking.

## Supplementary Note 2: Design and characterization of FPB metasurfaces

For simplicity in metasurface design and analysis, the cylindrical coordinate system is transformed into a Cartesian coordinate system. Any incident polarized beam on the metasurface can be decomposed into RCP and LCP spin eigenstates, whose Jones vectors are defined as:

$$\left|LCP\right\rangle = \begin{bmatrix} 1 \\ i \end{bmatrix}, \left|RCP\right\rangle = \begin{bmatrix} 1 \\ -i \end{bmatrix} \quad (6)$$

In this work, the corresponding phases are defined as $\varphi_L = 0$ and $\varphi_R = l\varphi$, where $\varphi = \arctan(y/x)$ denotes the azimuthal angle. The topological charge $l$ is set to +1 for Meta1 and -1 for Meta2. The metasurface's polarization transformation is characterized by the spatially varying Jones matrix $J(x,y)$, which satisfies the following eigen-equation for the spin basis:

$$J(x,y)\left|LCP\right\rangle = e^{i\varphi_L}\left|RCP\right\rangle, J(x,y)\left|RCP\right\rangle = e^{i\varphi_R}\left|LCP\right\rangle \quad (7)$$

Solving this yields the analytical expression for the metasurface Jones matrix:

$$J(x,y) = \begin{bmatrix} \dfrac{\exp\left[i\varphi_L(x,y)\right] + \exp\left[i\varphi_R(x,y)\right]}{2} & \dfrac{i\exp\left[i\varphi_R(x,y)\right] - i\exp\left[i\varphi_L(x,y)\right]}{2} \\ \dfrac{i\exp\left[i\varphi_R(x,y)\right] - i\exp\left[i\varphi_L(x,y)\right]}{2} & \dfrac{-\exp\left[i\varphi_L(x,y)\right] - \exp\left[i\varphi_R(x,y)\right]}{2} \end{bmatrix} \quad (8)$$

Under the constraints of symmetry and unitarity, the spatially varying Jones matrix $J(x,y)$ can be expressed in the canonical form $J(x,y) = Q\Lambda Q^{-1}$, where $Q$ is a real unitary matrix describing the rotation of the local birefringent axis, and $\Lambda$ is a diagonal matrix encoding the phase delays along the principal optical axes[50]. For birefringent nanostructures constituting the metasurface, the diagonal elements of $\Lambda$ correspond to the phase shifts ($\mathrm{Phase}_x$, $\mathrm{Phase}_y$) imparted along the orthogonal fast and slow axes, while $Q$ defines the in-plane rotation angle $\alpha$ of the local optical axis with respect to the global $x$-$y$ coordinate frame. These parameters, $\mathrm{Phase}_x$, $\mathrm{Phase}_y$, and $\alpha$, jointly govern the total phase modulation imparted by each meta-atom, where $\mathrm{Phase}_x$ and $\mathrm{Phase}_y$ determine the propagation phase, and the rotation angle $\alpha$ contributes to the geometric phase. For the given spin-multiplexed phase $\varphi_L$ and $\varphi_R$, the required phase shifts as well as the rotation angle are derived from the following equations:

$$Phase_x = \frac{\varphi_L + \varphi_R}{2}, \quad Phase_y = \frac{\varphi_L + \varphi_R}{2} - \pi \tag{9}$$

$$\alpha = \frac{\varphi_L - \varphi_R}{4} \tag{10}$$

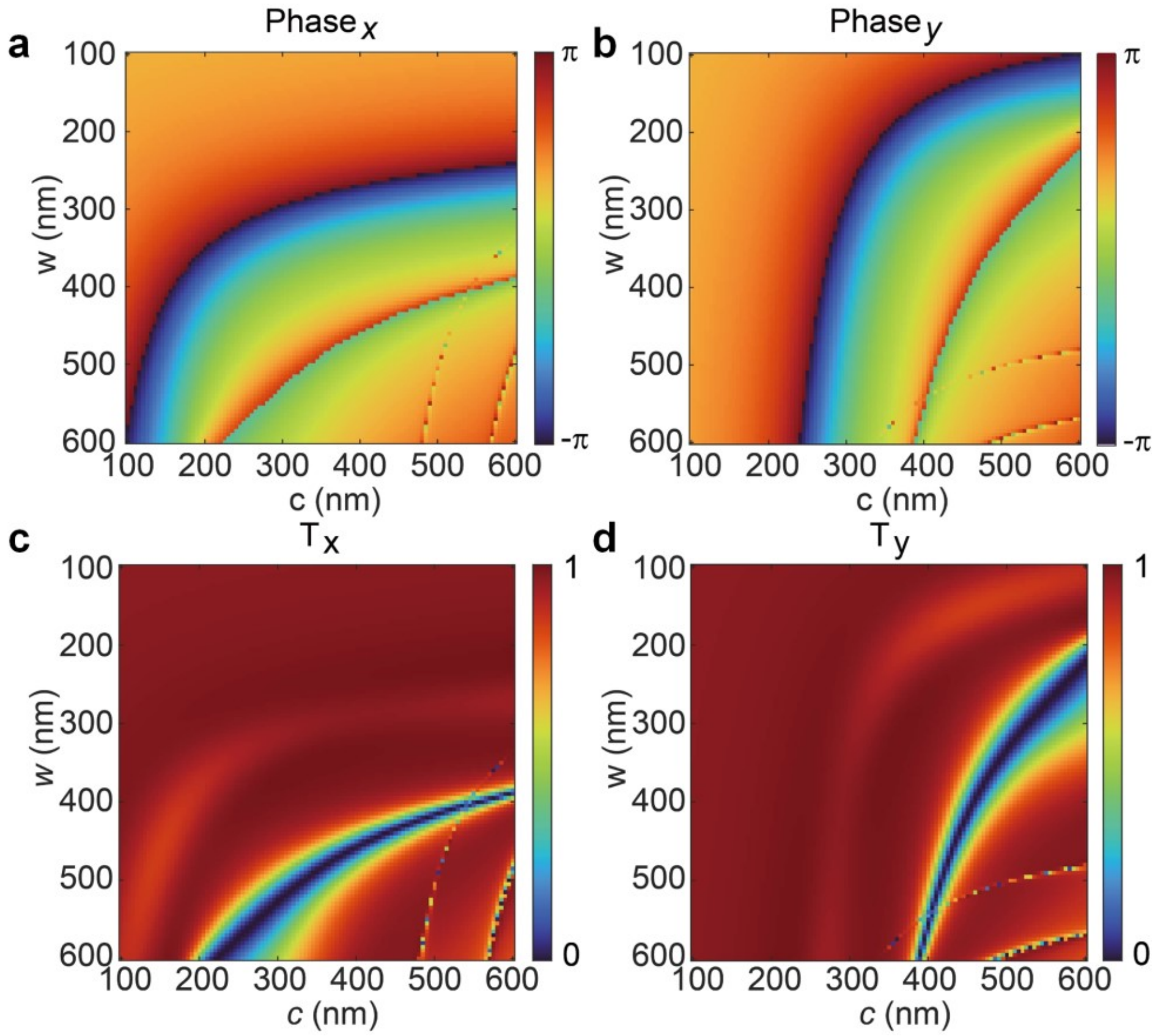

**Fig. S2. Simulation results of the FPB metasurface.** (a, b) Simulated phase shifts for $x$-polarized ($Phase_x$) and $y$-polarized ($Phase_y$) light, respectively, as functions of the meta-atom's geometric parameters ( width $w$ and length $c$ ) at a wavelength of 1550 nm. (c, d) Corresponding simulated power transmission coefficients for $x$-polarized ($T_x$) and $y$-polarized ($T_y$) light.

To ensure high transmittance at the telecom wavelength of 1550 nm, we utilize silicon nanopillars as the constituent elements of the metasurface. The metasurface is fabricated on a sapphire substrate, with a lattice period $P = 700$ nm (approximately half the operating wavelength) and a fixed nanopillar height $h = 800$ nm. The desired phase modulation is achieved by tuning the in-plane geometric parameters, nanopillar width $w$, length $c$, and orientation angle $\alpha$. A rigorous coupled-wave analysis simulation is conducted to determine the phase response and transmittance over a parameter sweep of $w$, $c \in [100, 600]$ nm, as shown in Fig. S2. Based on the simulation results, nanostructure geometries with optimal transmittance and target phase coverage are selected to realize two 5 mm-diameter FPB metasurfaces tailored for generating and demodulating FPBs.

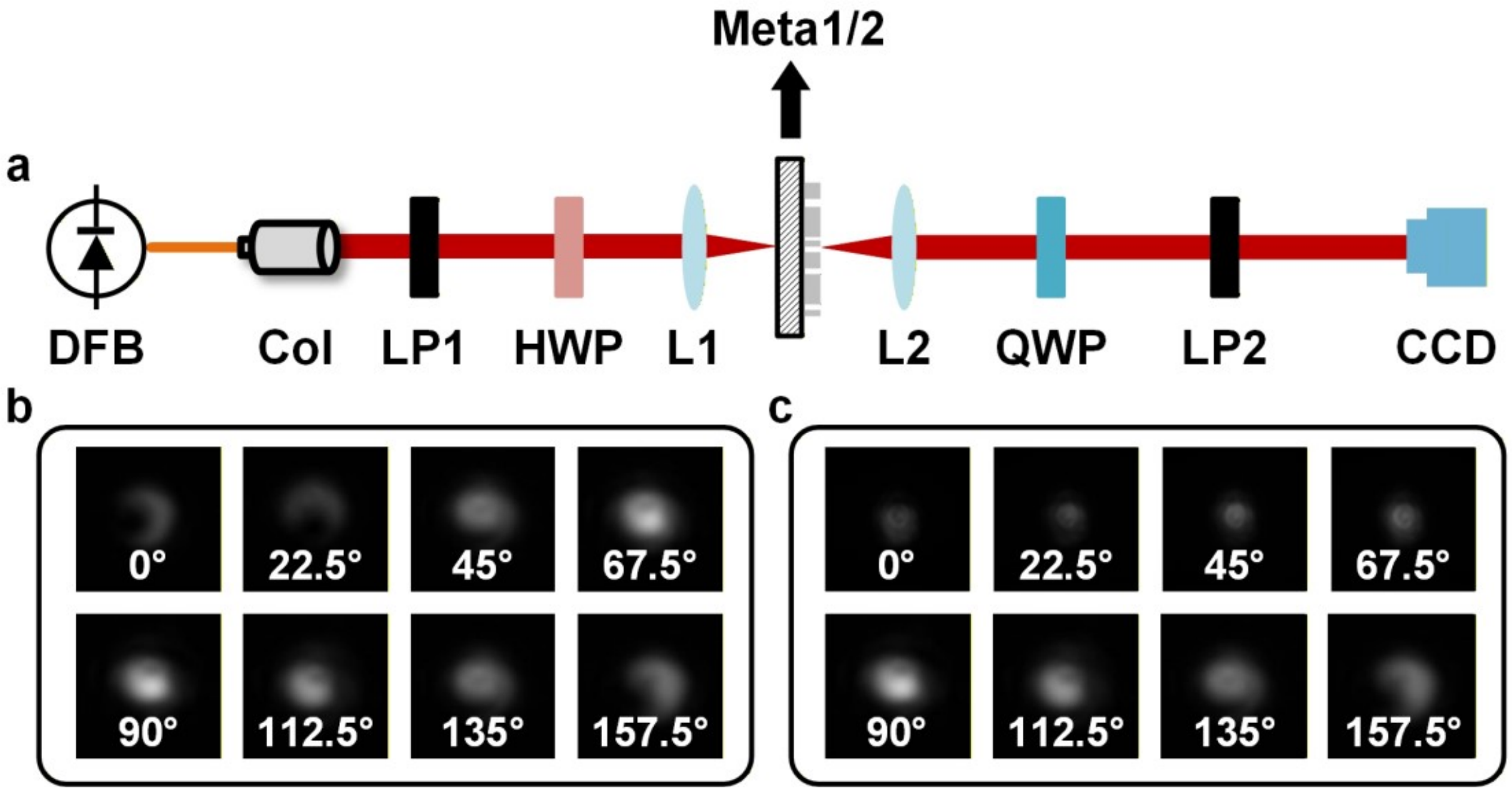

**Fig. S3. Characterization of FPB metasurface performance.** (a) Experimental optical setup for Stokes parameter measurement based on the QWP rotation method. DFB: distributed feedback semiconductor laser; Col: fiber collimator; LP: linear polarizer; HWP: half-wave plate; L: lens; QWP: quarter-wave plate; CCD: optical camera. (b) Measured light intensity distributions at eight QWP rotation angles for Meta1. (c) Corresponding intensity distributions for Meta2 under the same rotation conditions.

The Stokes parameters of the modulated optical field are measured using the quarter-wave plate (QWP) rotation method to evaluate both the transmissivity and polarization conversion efficiency (PCE) of the fabricated metasurfaces. The experimental setup is illustrated in Fig. S3(a). Due to the birefringent nature of the sapphire substrate, the metasurface samples are carefully aligned with their optical axis in the horizontal plane. A distributed feedback laser is coupled into free space via a fiber collimator, and the polarization state of the incident beam is adjusted to linear *x*-polarization using a linear polarizer and a half-wave plate (HWP). After transmission through the metasurface, the beam is collimated by lens L2 and directed through a rotating QWP, which is incrementally adjusted from 0° to 180° in 22.5° steps. The resulting light intensity is recorded by a CCD camera. Figure S3 (b, c) presents the corresponding intensity distributions for Meta1 and Meta2 across the eight QWP rotation angles.

The transmissivity T of the metasurface is defined as:

$$Trans. = \frac{I_t}{I_0} \times 100\%, \tag{11}$$

where $I_t$ denotes the beam energy measured after transmission through the metasurface and lens L2, and $I_0$ is the incident beam energy measured just after the HWP, prior to the metasurface. The PCE of the metasurface quantifies the proportion of the transmitted beam converted into the RCP component—i.e., the vortex-modulated portion of the beam. The energy of the RCP component is given by $I_{RCP} = 1/2(S_0 + S_3)$, where $S_0$ and $S_3$ are Stokes parameters. The PCE is then calculated as:

$$PCE = \frac{I_{RCP}}{0.5 I_0} \times 100\% = \left(1 + \frac{S_3}{S_0}\right) \times 100\% \tag{12}$$

**Supplementary Figure S4**

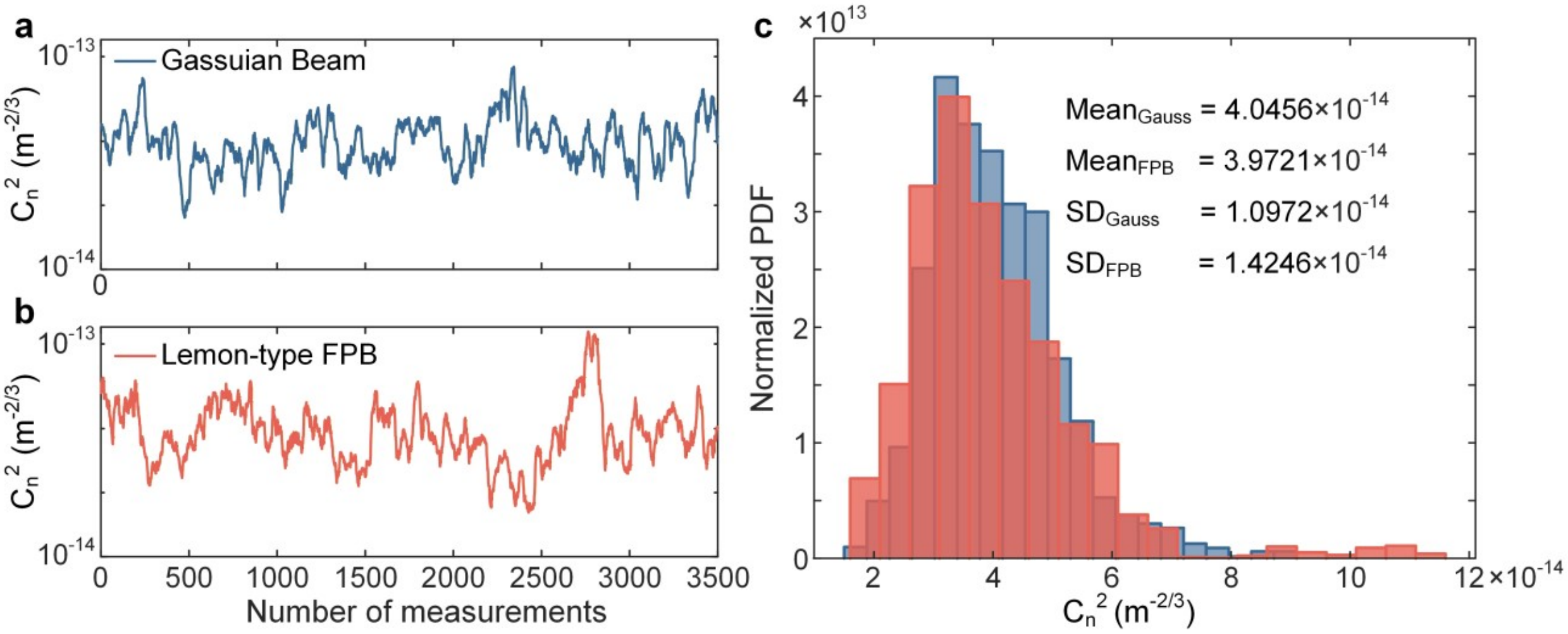

**Fig. S4. Real-world turbulence strength monitored by DIMM.** (a, b) Measured atmospheric turbulence refractive index structure parameters in a 3.2 km FSO link for the Gaussian Beam (a) and lemon-type FPB (b) under 3500 seconds, respectively. (C) Histograms of the probability density distribution function of refractive index structure parameters for both beams.

## Supplementary Note 3: BER performance as a function of received optical power under realistic atmospheric turbulence

To evaluate the system performance of the CTFA security scheme and the proposed FPB metasurfaces, we measured the BER as a function of received optical power (ROP) under different scenarios using a 30 Gbaud QPSK signal at a center wavelength of 1550 nm. Figure S5 summarizes the BER performance for both back-to-back (btb) and 3.2 km field transmission conditions. In the btb case, the integration of the CTFA encryption module introduces a modest power penalty of approximately 0.5 dB relative to the unencrypted baseline. Under real-world turbulence conditions over a 3.2 km horizontal free-space optical (FSO) link, receiver sensitivity is markedly degraded due to atmospheric effects. Without FPB metasurface assistance—that is, when transmitting a linearly polarized Gaussian beam—the ROP required to reach the HD-FEC threshold increases to approximately –26 dBm, corresponding to a total power penalty of ~4.5 dB. However, with the application of FPB metasurfaces for turbulence mitigation, this penalty is reduced to ~2.7 dB. Across the full ROP range, the BER achieved with FPB consistently outperforms that of the Gaussian beam, with particularly pronounced improvements observed at higher ROP levels.

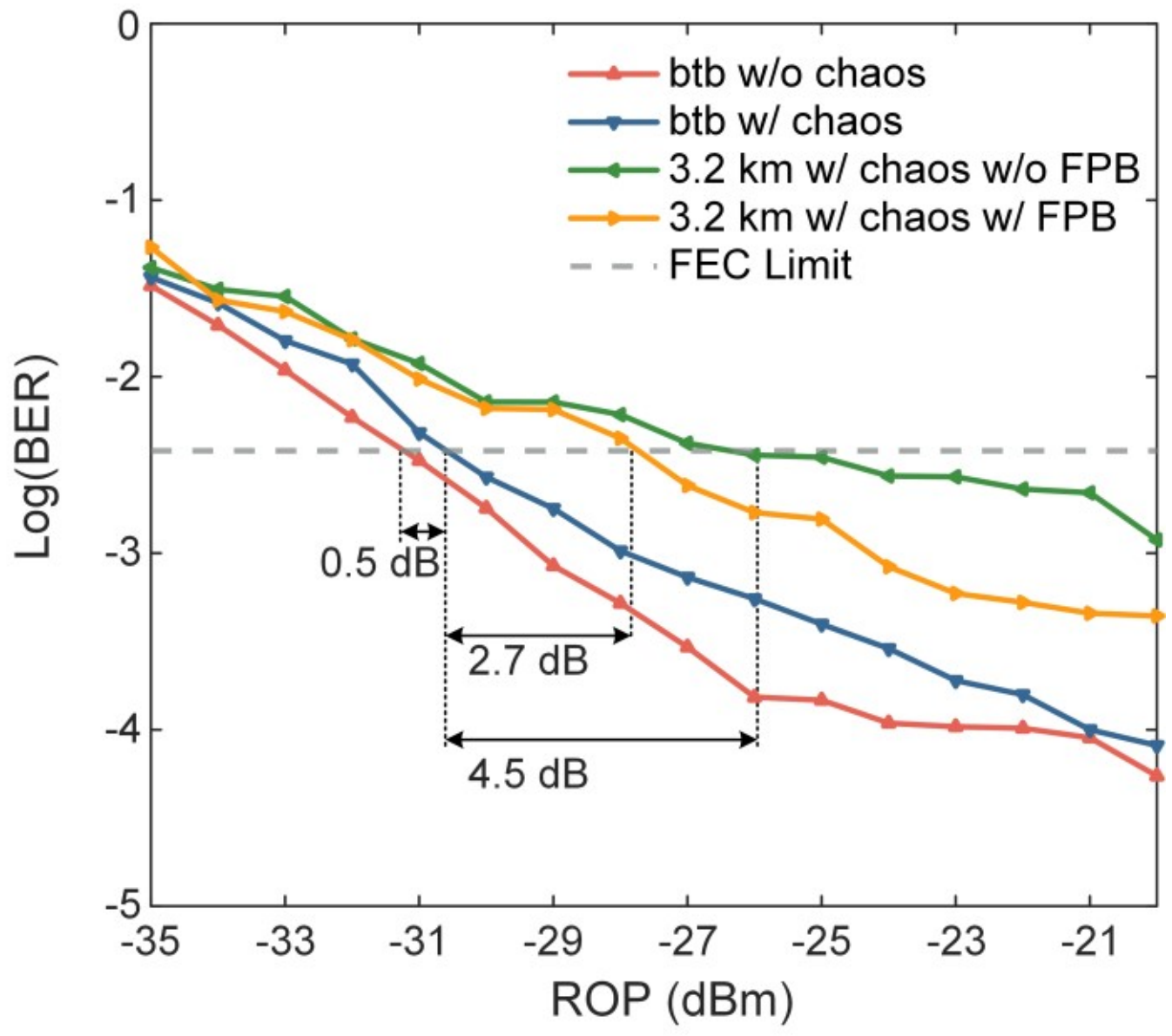

**Fig. S5.** BER as a function of the received optical power for the communication system. The BER performance under the cases of btb and 3.2 km transmission is shown in the figure. The BER value for each received optical power is calculated as the average of ten consecutive measurements.

## Supplementary Note 4: Self-homodyne phase autocorrelation time-delay detection method

The "just-in-time" descrambling of the chaotic phase is crucial for accurate decryption. Since the encrypted signals and the common chaotic signal share the same physical link (optical fiber or free-space link) after WDM, the primary source of time delay between chaotic phase scrambling and descrambling arises from hardware mismatches between the CTFA and CTFDA modules, as well as delays introduced by CFBG2 and CFBG3. Here we propose a method, named self-homodyne phase autocorrelation time-delay detection, to measure the time delay mismatch, for subsequent optical path compensation. By coherently detecting the optical signal whose phase is modulated by chaotic driving signals $I_A$ and $I_B$ in succession, the time delay mismatch can be obtained from the autocorrelation of the optical signal's phase. The schematic diagram for time delay mismatch measurement is shown in Fig. S6.

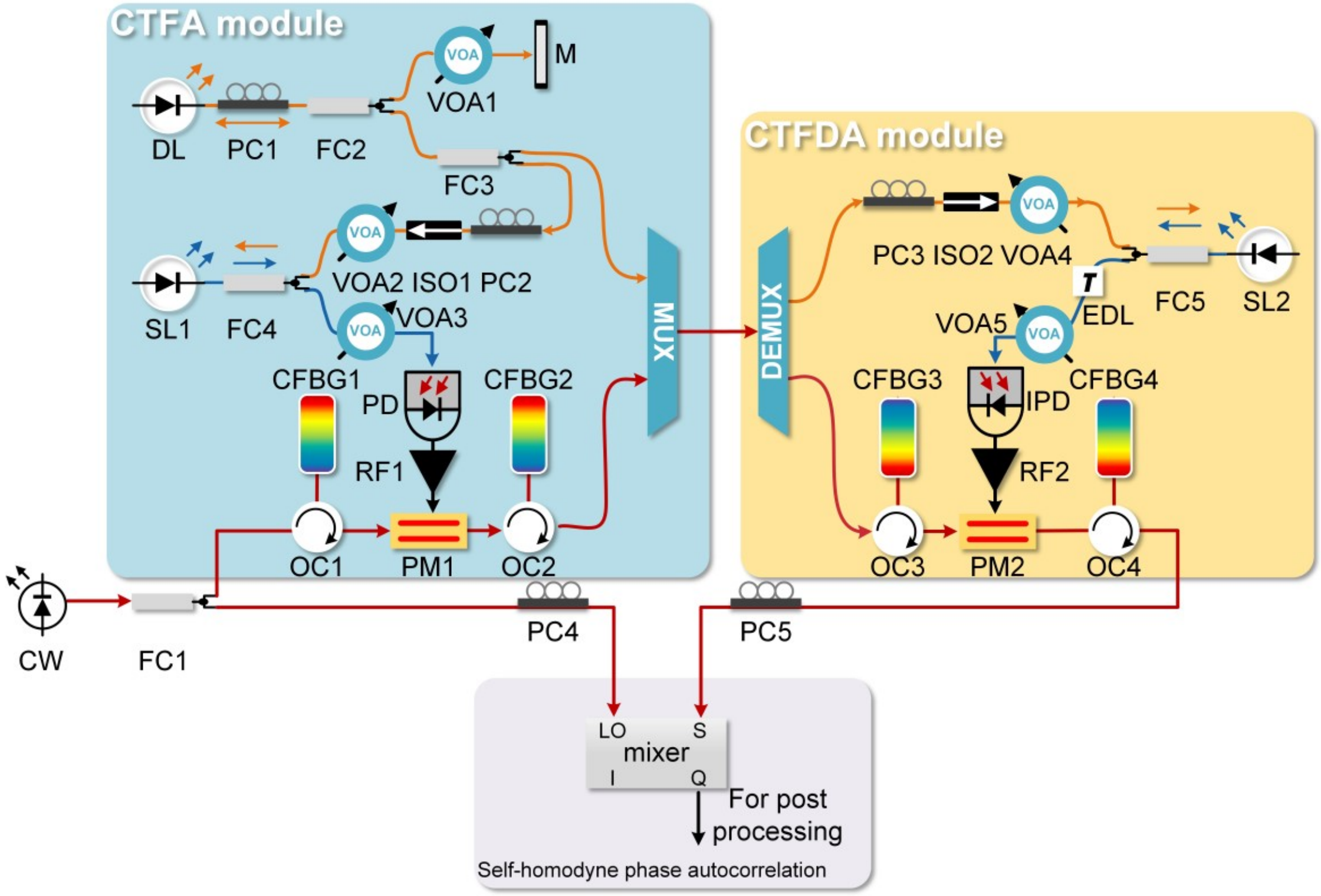

**Fig. S6. Schematic diagram for time delay mismatch measurement.** DL, drive laser; PC, polarization controller; FC, fiber coupler; VOA, variable optical attenuator; M, fiber mirror; ISO, optical isolator; SL, slave

laser; (I)PD, (inverse)photodetector; RF, radio-frequency amplifier; OC, optical circulator; CFBG, chirped fiber Bragg grating; PM, electro-optical phase modulator; EDL, electrically controlled fiber delay line.

The optical field $E_{CW}$ of a continuous-wave (CW) laser can be written as:

$$E_{CW}(t) = A_{CW}\exp\left[i\left(\omega_0 t + \phi_0 + \phi_N(t)\right)\right], \tag{13}$$

where $A_{CW}$, $\omega_0$, $\phi_0$, and $\phi_N(t)$ are the amplitude, the angular frequency, the initial phase, and the phase noise of the optical field, respectively. After being modulated by phase modulators PM1 and PM2 in succession, the optical field $E_{PM}$ can be written as:

$$E_{PM}(t) = A_{PM}\exp\left[i\left(\omega_0 t + \phi_0 + \frac{C_A(t-\Delta t)}{V_\pi}\pi - \frac{C_B(t)}{V_\pi}\pi + \phi_N(t)\right)\right], \tag{14}$$

where $C_A(t)$ and $C_B(t)$ are the electrical chaotic driving signals $I_A$ and $I_B$, respectively. $\Delta t$ represents the time delay mismatch between them and $V_\pi$ is the half-wave voltage of the phase modulators. Mixing the optical signal $E_{PM}$ with $E_{CW}$, the mixer outputs the quadrature term as

$$S(t) = R_d\left|iE_{PM}(t) + E_{CW}(t)\right|^2 = R_d\left(A_{CW}^{\ 2} + A_{PM}^{\ 2} + 2A_{CW}A_{PM}\cos\left(\Delta\phi(t) + \frac{\pi}{2}\right)\right), \tag{15}$$

with $\Delta\phi(t) = \left[\frac{C_A(t-\Delta t)}{V_\pi}\pi - \frac{C_B(t)}{V_\pi}\pi\right]$ for the simplification purpose. $R_d$ is the responsivity of the mixer's photo-detectors.

Ignoring the direct-current (DC) terms of the Eq. (15), one can obtain the difference frequency component $U(t)$

$$U(t) = \frac{S(t) - R_d\left(A_{CW}^{\ 2} + A_{PM}^{\ 2}\right)}{2R_d A_{CW} A_{PM}} = \cos\left(\Delta\phi(t) + \frac{\pi}{2}\right). \tag{16}$$

And one can further get

$$\Delta\phi(t) = a\cos\left(U(t)\right) - \frac{\pi}{2}, \text{ s.t. } \Delta\phi(t) \in \left[-\frac{\pi}{2}, \frac{\pi}{2}\right]. \tag{17}$$

The limited condition of Eq. (17) can be easily satisfied by adjusting the signal strengths of $C_A(t)$ and $C_B(t)$. Thanks to the high correlation between these two signals, a pair of characteristic peaks symmetrically distributed relative to $t = 0$ can be observed in the autocorrelation curve of $\Delta\phi(t)$. Note that the peak should appear at the valley position of the autocorrelation curve, as $C_A(t)$ and $C_B(t)$ are inverting.

The measurement results of the time delay mismatch from our experiments are shown in Fig. S7. Initially, the characteristic peaks are found at ±10.110 ns (Fig. S7(a)), when the time delay of an electrically controlled fiber delay line (EDL) is set to 0 ps. Next, with the time delay adjusted to 585 ps, the corresponding peaks shifted to ±9.525 ns (Fig. S7(b)). The measured values align with the set value, and it suggests that the optical path containing the EDL should be further extended to achieve precise time delay matching. A fiber with a length of about 2 m is added to the corresponding optical path, and the measurements are repeated. The final results are shown in Fig. S7(c) (time delay of EDL set as 0 ps) and Fig. S7(d) (time delay of EDL set as 700 ps), confirming that the time delay mismatch falls within the operational range of the EDL (0~700 ps).

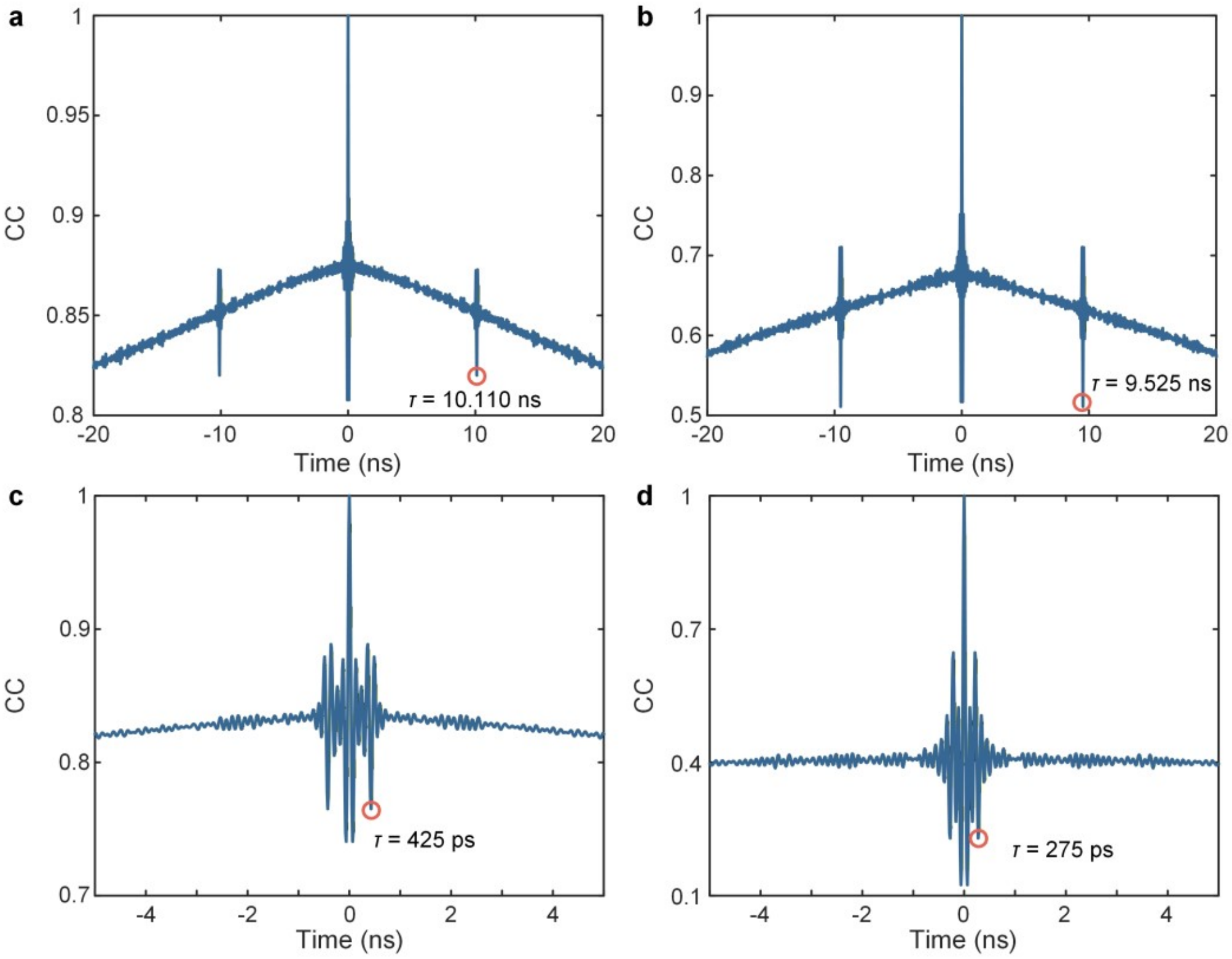

**Fig. S7. Time delay mismatch measurements for optical path compensation.** (a) Characteristic peaks are observed at $\tau = \pm 10.110$ ns with the EDL set to 0 ps. (b) Characteristic peaks are observed at $\tau = \pm 9.525$ ns with the EDL set to 585 ps. (c) After adding a 2 m optical fiber, characteristic peaks are observed at $\tau = \pm 425$ ps with the EDL set to 0 ps. (d) After adding a 2 m optical fiber, characteristic peaks are observed at $\tau = \pm 275$ ps with the EDL set to 700 ps.

Then, the bit error ratio (BER) of the decrypted signal (30 Gbaud QPSK) is swept by stepwise setting the time delay of the EDL, and the corresponding result is shown in Fig. S8. When the EDL is set to around 425 ps, a minimal BER of $1.59\times10^{-4}$ is found, which is in coincidence with the results of the time delay mismatch measurements performed previously (Fig. S7(c)).

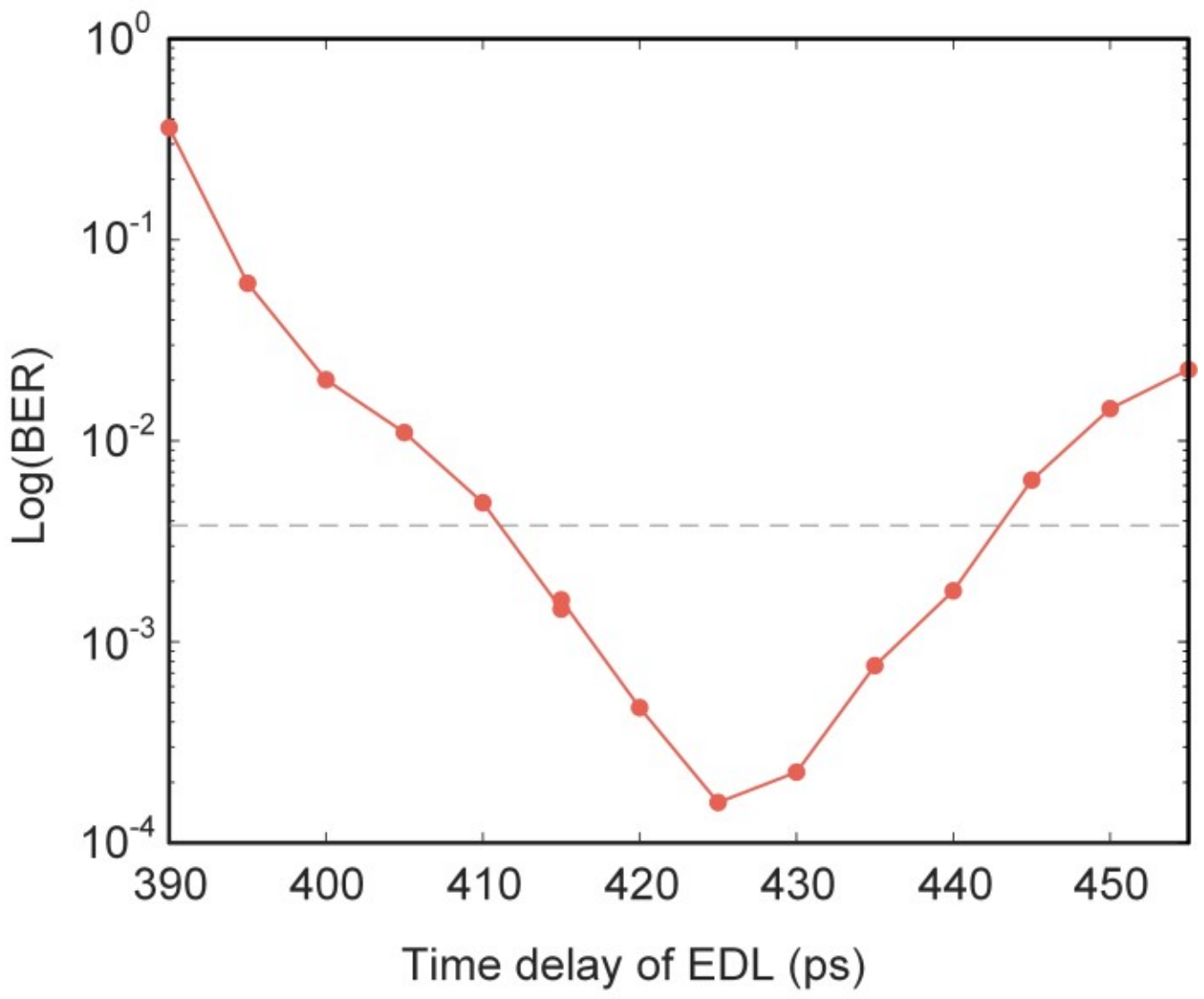

**Fig. S8. Measured BER at different time delays of EDL.** The experimental data are shown in red solid dots, and the HD-FEC threshold is indicated by the gray dashed line. A minimal BER of $1.59 \times 10^{-4}$ is found at around 425 ps (EDL's set value), in coincidence with the time delay mismatch measurements.

## Supplementary Figure S9

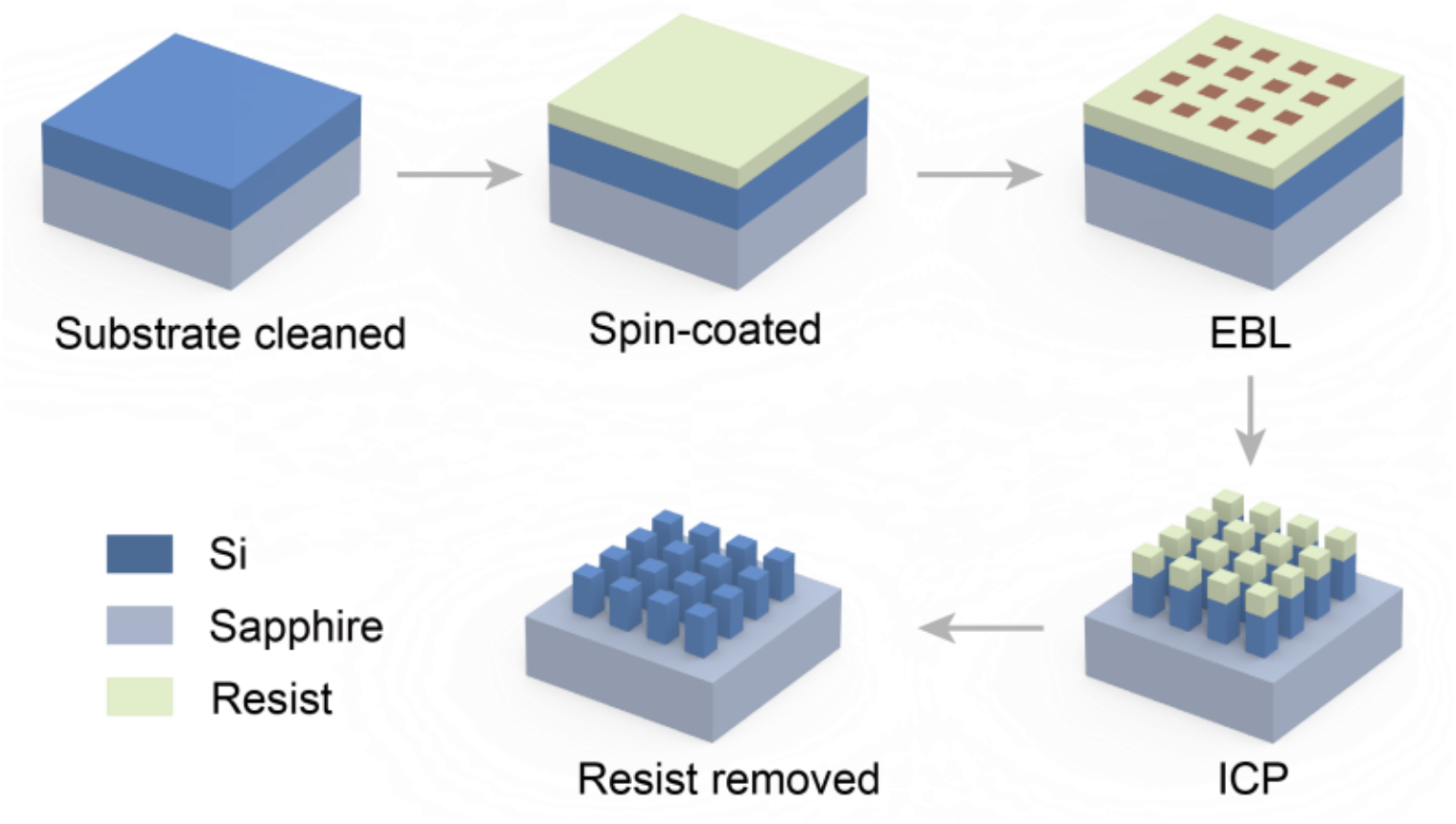

**Fig. S9. Schematic of the metasurface fabrication process.**